\documentstyle[twocolumn,seceq,epsf]{jpsj}

\def\runtitle{Electronic structures of MgB{$_2$} under uniaxial compression}
\def\runauthor{Kazuaki {\sc Kobayashi} and Kazuo {\sc Yamamoto}}

\title
{
Electronic Structures of MgB{$_2$} under Uniaxial and Hydrostatic
Compression
}

\author
{ 
Kazuaki {\sc Kobayashi}\footnote{Present author's (K. K.) URL address:\newline
http://www.geocities.co.jp/Technopolis/4765/index.html}  and Kazuo
{\sc Yamamoto}$^{1,}$
}

\inst
{
Advanced Materials
Laboratory, National Institute for Materials Science, Namiki 1-1,
Tsukuba-shi, Ibaraki 305-0044\\
$^1$Kanagawa Institute of Technology, 1030 Shimo-ogino, Atsugi,
Kanagawa 243-0292
}

\recdate
{
\today
}

\abst
{
Electronic and lattice properties of MgB{$_2$} under uniaxial and
hydrostatic compression are calculated.
Lattice properties are optimized automatically by using the
first-principles molecular dynamics (FPMD) method.
Features of the electronic band structures under uniaxial and
hydrostatic compression are quite different each other.
}

\kword
{
MgB{$_2$}, uniaxial compression, high pressure, first-principles,
electronic structure, phonon frequency
}

\begin{document}
\sloppy
\maketitle

\section{Introduction}
 Very recently, a new high {\it T}{$_c$} intermetallic superconductor
material MgB{$_2$} ({\it T}{$_c$} = 39 K) was discovered by
J. Akimitsu group~\cite{rf:1}.
 MgB{$_2$} has a AlB{$_2$} crystal symmetry(P6/mmm) and the number of
atoms in a unit cell is three.
 Already, a variety of extended
experimental~\cite{rf:exp1,rf:exp2,rf:exp3,rf:exp4,rf:exp5} and
theoretical~\cite{rf:th1,rf:th2,rf:th3,rf:th4,rf:th5,rf:th6,rf:Loa,rf:Wan}
studies have been devoted~\cite{rf:HTC}.
 The electronic structures of MgB{$_2$} under high pressure
(hydrostatic)~\cite{rf:Loa} and with different lattice
constants~\cite{rf:Wan} are calculated by FLAPW, respectively.

 In this study, we calculate the electronic and lattice properties of
MgB{$_2$} under uniaxial compression by using the first-principles
molecular dynamics (FPMD) and to compare with results under hydrostatic
compression.
 It is possible to obtain the optimized lattice properties under high
pressure conditions using FPMD.
 We investigate the changes of electronic band structures under a
variety of compression (uniaxial, hydrostatic, varying the {\it c/a}
ratio).
 Phonon frequencies at the {$\Gamma$} point under uniaxial and hydrostatic 
compression are calculated in this study.

\section{Method of Calculation}
 The present calculation is based on the local density approximation in
the density functional theory~\cite{rf:HK,rf:KS} with the Wigner
interpolation formula~\cite{rf:Wigner} for the exchange-correlation.
 The optimized pseudopotential (Mg and B) by Troullier and Martins
(TM)~\cite{rf:TM} is used in order to reduce the number of plane waves.
 Non-local parts of pseudopotential are transformed to the
Kleinman-Bylander separable forms~\cite{rf:KB}.
 No ghost bands by this treatment are found in our calculation.
 A partial core correction (PCC)~\cite{rf:PCC} is considered for Mg
pseudopotential.
 The wave function is expanded in plane waves, and the energy cutoff
is 81 Ry with the maximum number of plane waves being about 2000.
 The number of sampling k-points is 95 in the 1/24 of the Brillouin
zone (BZ) for all calculated systems.
 In the cell optimization, the number of k-points is fixed because a
change of the {\it c/a} ratio under compression is not so large ({\it
c/a} = 0.868 {$\sim$} 1.123). 

 The outline of our calculational process with the cell optimization
is as follows.
 In this calculation, the system has kept to the AlB{$_2$} crystal
symmetry in the process of the cell optimization.
 Therefore, it does not discuss a structural phase transformation
under uniaxial and hydrostatic compression in this study.
 At first, an electronic part is quenched to the ground state of a
given initial lattice parameter, and optimization of electronic and
unit cell shape parts is carried out simultaneously in the next step.
 Stresses~\cite{rf:NM} acting on unit cell surfaces are calculated,
and the lattice parameters are tuned by using them.
 We can estimate that a numerical error of the calculated pressure is
less than about 1 GPa. 
 There is no effect to discuss qualitatively although this value (1
GPa) may be slightly large.

 We calculate in three conditions as follows.
 The first is the uniaxial compression. 
 The uniaxial compression acts along a direction of {\it c}-axis. 
 Values of external compression along {\it a}- and {\it b}-axis
are 0 GPa in this case.
 The second is the hydrostatic compression. 
 The third is to vary the {\it c/a} ratio when the values of {\it a} and
{\it b} are fixed
in the equilibrium lattice parameters.
 In the case of varying the {\it c/a} ratio, all surfaces of the unit
cell have non-zero internal pressures.

\section{Results and Discussion}
 The optimized lattice properties in this calculation are tabulated in
Table~\ref{table:1}.

\begin{table}[htbp]
\caption{\hspace{0.0cm} Equilibrium lattice constants[{\AA}] and the
{\it c/a} ratio of MgB{$_2$}. ``{\it P}z'' indicates the uniaxial
compression along {\it c}-axis. ``{\it P}'' indicates the hydrostatic
compression. ``{\it c/a}'' indicates the varying {\it c/a}. The value
of compression ``{\it c/a}'' indicates that along {\it c}-axis.}
\label{table:1}
\begin{tabular}{@{\hspace{\tabcolsep}\extracolsep{\fill}}cccc} \hline
            &  {\it c}    &  {\it a}    &  {\it c/a}  \\ \hline
 0 GPa      & 3.421       &  3.047      &   1.123     \\ \hline
 10 GPa({\it P}z) & 3.307       &  3.053      &   1.083     \\ \hline
 20 GPa({\it P}z) & 3.183       &  3.067      &   1.038     \\ \hline
 30 GPa({\it P}z) & 3.045       &  3.091      &   0.985     \\ \hline
 40 GPa({\it P}z) & 2.932       &  3.109      &   0.943     \\ \hline
 50 GPa({\it P}z) & 2.838       &  3.125      &   0.908     \\ \hline
 10 GPa({\it P})  & 3.324       &  2.998      &   1.109     \\ \hline
 20 GPa({\it P})  & 3.246       &  2.958      &   1.097     \\ \hline
 30 GPa({\it P})  & 3.179       &  2.923      &   1.088     \\ \hline
 40 GPa({\it P})  & 3.122       &  2.894      &   1.079     \\ \hline
 50 GPa({\it P})  & 3.072       &  2.868      &   1.071     \\ \hline
 22.3 GPa({\it c/a}) & 3.175    &  3.047    &   1.042 \\ \hline
 48.4 GPa({\it c/a}) & 2.910    &  3.047    &   0.955 \\ \hline
 84.6 GPa({\it c/a}) & 2.646    &  3.047    &   0.868 \\ \hline
 Exp~\cite{rf:1} & 3.524       &  3.086      & 1.142  \\ \hline
\end{tabular}
\end{table}

 Calculated bulk modulus of MgB{$_2$} is 145 GPa, which agrees with
other theoretical result~\cite{rf:Loa}.
 A heat of formation of MgB{$_2$} is 0.61 eV/cell.
 The calculated equilibrium lattice constant of {\it c}-axis is shorter by 3
\% than that of experiment~\cite{rf:1}.
 The number of sampling k-points (= 95) may be insufficient for
discussing quantitatively the electronic and lattice properties
because MgB{$_2$} is metallic.
 In addition, this number is not adjusted in varying the unit cell
volume.
 There is no problem for discussing qualitatively the electronic band
structure changes under uniaxial and hydrostatic compression in this
study.
 To check the convergence, we also calculate in the condition of the
259 k-points and 100 Ry under 0 GPa.
 A deviation of the equilibrium lattice constants is less than about
0.01 {\AA}, and slightly improve the error of the lattice constant of
{\it c}-axis from 3 \% to 2.6 \%.

 The electronic band structures of the optimized lattice parameters
are calculated.
 The electronic band structures under uniaxial compression (0, 20, 50
GPa) are shown in Fig.~\ref{fig:1}.
 The electronic band structures under hydrostatic compression (20,
50 GPa) are also shown in Fig.~\ref{fig:1}.
 The electronic band structure ({\it c/a} = 0.955, {\it P}z = 48.4 GPa) in
varying the {\it c/a} ratio is also shown in Fig.~\ref{fig:1}.
 As a result, the electronic band structures at 20, 50 GPa under
hydrostatic compression are similar to that at 0 GPa.
 The small change of the {\it c/a} ratio under hydrostatic compression
indicates the isotropic compressibility.
 This trend agrees with the other theoretical study~\cite{rf:Loa}
although the {\it c/a} ratio (= 1.123, P = 0 GPa) at present study is
slightly different from the other theoretical result (=
1.1468)~\cite{rf:Loa}.
 The widths of band at 20, 50 GPa is larger than that at 0
GPa (that of 50 GPa is largest) as shown in Table~\ref{table:2}.
 On the other hand, the electronic structures under uniaxial
compression are quite different from those under hydrostatic
compression at the Fermi level.
 The {$\sigma$ bands at the {$\Gamma$}- {\it A}} line shift to lower
energy as the pressure increase. 
 The electronic structure with varying the {\it c/a} ratio ({\it c/a} =
0.955, {\it P}z = 48.4 GPa) is also different from those under hydrostatic
compression as shown in Fig.~\ref{fig:1}.
 This band structure (a = a{$_0$} and c = 0.85c{$_0$}, a{$_0$} and
c{$_o$} are the calculated equilibrium lattice constant in this
study.) is corresponding to that of a = a{$_0$} and c =
0.8c{$_0$}~\cite{rf:Wan} (a{$_0$} and c{$_0$}~\cite{rf:Jones}) and
they are similar to each other.

 We calculate density of states (DOS) at 0 GPa, 50 GPa(hydrostatic),
50 GPa(uniaxial), nearly 50 GPa(varying {\it c/a}, {\it P}z = 48.4
GPa).
 The density of states at Fermi level (N({$\epsilon_f$})) at 0 GPa,
50 GPa(hydrostatic), 50 GPa(uniaxial), nearly 50 GPa(varying {\it
c/a}) are 1.0, 0.94, 0.71, 0.77, respectively, where the values of
N({$\epsilon_f$}) are normalized as 1.0 for N({$\epsilon_f$}) at 0 GPa. 
 The number of k-points  is not enough to discuss quantitatively 
the values of N({$\epsilon_f$}).
 The values of N({$\epsilon_f$}) under compression are smaller than
that under 0 GPa and the decrease of N({$\epsilon_f$}) under
anisotropic compression is larger than that under hydrostatic
(isotropic) compression.
 The density of states at 0 GPa and 50 GPa(hydrostatic) are similar to
each other except for the band width.
 The density of states at 50 GPa (uniaxial) and nearly 50 GPa(varying
{\it c/a}) are quite different from those at 0 GPa and 50
GPa(hydrostatic).

\begin{table}[htbp]
\caption{\hspace{0.0cm} The values of the band width from the bottom
of the valence band to the Fermi level. ``{\it P}z'' indicates the uniaxial
compression along {\it c}-axis. ``{\it P}'' indicates the hydrostatic
compression. ``{\it c/a}'' indicates the varying {\it c/a}.}
\label{table:2}
\begin{tabular}{@{\hspace{\tabcolsep}\extracolsep{\fill}}ccc} \hline
            &  Width(eV)  &  {\it c/a}    \\ \hline
 0 GPa      &  12.6       &  1.123  \\ \hline
 20 GPa({\it P}z) &  12.9       &  1.038  \\ \hline
 50 GPa({\it P}z) &  14.0       &  0.908  \\ \hline
 20 GPa({\it P})  &  13.3       &  1.097  \\ \hline
 50 GPa({\it P})  &  14.0       &  1.071  \\ \hline
 48.4 GPa({\it c/a}) & 13.9   &  0.955  \\ \hline
\end{tabular}
\end{table}

 We calculate phonon frequencies {\it f}{$_{\rm w}$} at the {$\Gamma$}
point under 0 GPa, 50 GPa(uniaxial) and 50 GPa(hydrostatic)
compression as shown in Table~\ref{table:3}.
 The calculated displacements of atoms in the unit cell are shown in
Fig.~\ref{fig:2}.
 The number of k-points is 512 (8{$\times$}8{$\times$}8) in the whole
Brillouin zone (BZ) in order to obtain the accurate phonon frequencies
because this calculation does not consider symmetry due to the
displacement of atoms in the unit cell.
 The values of the phonon frequencies at the {$\Gamma$} point are in
good agreement with other theoretical
results~\cite{rf:th1,rf:th3,rf:th6} except for the E{$_{2g}$} mode.
 The discrepancies of the phonon frequencies at the E{$_{1u}$},
A{$_{2u}$}, B{$_{1g}$} modes at present and the other theoretical
studies~\cite{rf:th1,rf:th3,rf:th6} are very small.
 On the other hand, the values of the E{$_{2g}$} mode (629[Present],
470~\cite{rf:th1}, 585~\cite{rf:th6}, 665~\cite{rf:th3} cm{$^{-1}$})
in the theoretical calculations has a large difference.
 The values of phonon frequencies under compression are
larger than that under 0 GPa because the cell volume is shrunk by
compression.
 Values of the volume ratio (V/V{$_0$}, V{$_0$}:equilibrium volume)
are 0.795 ({\it P} = 50 GPa) and 0.872 ({\it P}z = 50 GPa),
respectively.
 The deviations of lattice changes are -5.9 \% ({\it P} = 50 GPa),
+2.6 \% ({\it P}z = 50 GPa) along {\it a}-axis and -17.0 \% ({\it P} = 
50 GPa), -10.2 \% ({\it P}z = 50 GPa). 
 The value of the phonon frequency at the E{$_{2g}$} mode under
hydrostatic compression (50 GPa) is 886 cm{$^{-1}$}, in which the
increase of the phonon frequency is largest in all calculated phonon
frequencies.
 In particular, the change of the phonon frequency at the E{$_{2g}$}
mode under uniaxial compression ({\it P}z = 50 GPa) is largest in the
four modes in spite of the elongation of a B-B distance on the B
layer.
 The {$\sigma$} bands at the {$\Gamma$} - {\it A} line are completely
occupied by uniaxial compression.
 This occupation means that the number of electrons with regard to
{$\sigma$} bonding in the B layer increases.
 Increasing {$\sigma$} bonding electrons under uniaxial compression is
to strengthen the B-B bond, which leads to increase the value of the
phonon frequency at E{$_{2g}$} mode.
 On the other hand, the lattice change (-5.9 \%) under hydrostatic
compression (P = 50 GPa) directly strengthens the B-B bonding although
the {$\sigma$} bands at the {$\Gamma$} - {\it A} line are unoccupied.

\begin{table}[htbp]
\caption{\hspace{0.0cm} The values of the phonon frequencies {\it
f}{$_{\rm w}$} (cm{$^{-1}$}) at the {$\Gamma$} point. ``{\it P}z''
indicates the uniaxial compression along {\it c}-axis. ``{\it P}''
indicates the hydrostatic compression.}
\label{table:3}
\begin{tabular}{@{\hspace{\tabcolsep}\extracolsep{\fill}}cccc} \hline
            &  {\it f}{$_{\rm w}$} (0 GPa) &  {\it f}{$_{\rm w}$}
({\it P} = 50 GPa) & {\it f}{$_{\rm w}$} ({\it P}z =
50 GPa)   \\ \hline 
 E{$_{1u}$}   &  348   & 500  &  513     \\ \hline
 A{$_{2u}$}   &  398   & 594  &  517     \\ \hline
 E{$_{2g}$}   &  629   & 886  &  851     \\ \hline
 B{$_{1g}$}   &  707   & 814  &  747     \\ \hline
\end{tabular}
\end{table}

\section{Summary}
 We calculate the electronic and lattice properties of MgB{$_2$} under
uniaxial, hydrostatic compression and varying the {\it c/a} ratio by
using FPMD.
 The phonon frequencies at the {$\Gamma$} point under 0 GPa, 50
GPa(uniaxial) and 50 GPa(hydrostatic) are calculated.
 It is possible to explain the increasing the phonon frequencies under
compression by the analysis of the electronic band structures.
 Although the width of the electronic band structure is spread by
compression, the band structure change is not so large under
hydrostatic compression. 
 The electronic band structure changes ({$\sigma$} bands) are quite
large under uniaxial compression and varying the {\it c/a} ratio at
the Fermi level.
 We think that it is very important to investigate the difference of
the electronic band structures ({$\sigma$} bands) at the Fermi level
under hydrostatic, uniaxial and varying the {\it c/a} ratio.

 More accurate study (the number of k-points, possibility of a
pressure induced phase transition, etc.) is a near future
task. 


\begin{fullfigure}
\hfil\epsfxsize=16cm\epsfbox{MgB2_0_20_50.eps}\hfil
\caption{Energy band structures of MgB{$_2$} under uniaxial and
hydrostatic compression, {\it c/a} = 0.955 (varying the {\it c/a}
ratio). The Fermi level is indicated by the horizontal line.}
\label{fig:1}
\end{fullfigure}

\begin{fullfigure}
\hfil\epsfxsize=18cm\epsfbox{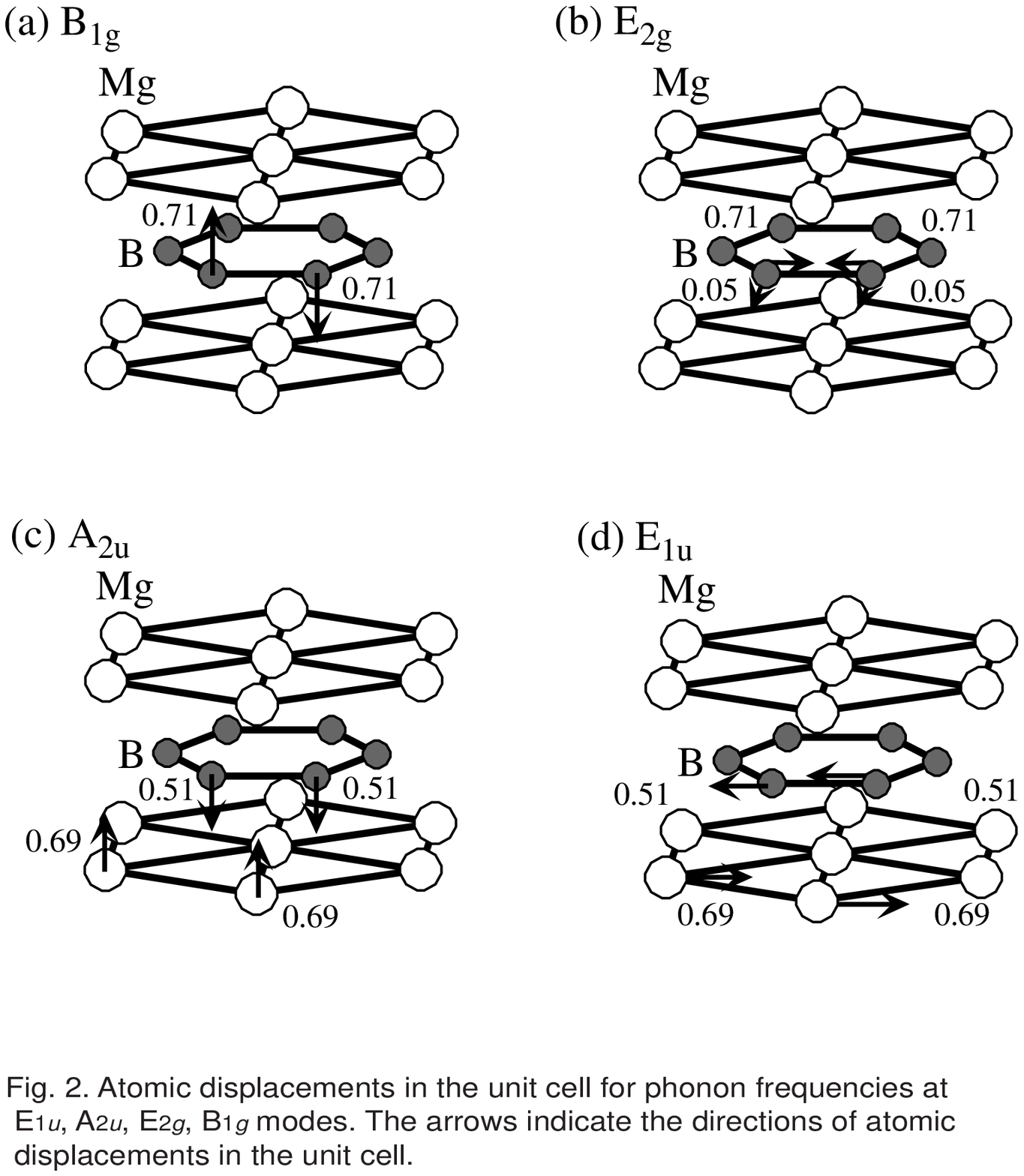}\hfil
\label{fig:2}
\end{fullfigure}

\section*{Acknowledgements}
We would like to thank Dr. M. Arai, Dr. K. Takemura, Dr. S. Higai and
Prof. S. Suzuki for valuable discussions.
  The numerical calculations were performed at The National Institute
for Research in Inorganic materials (AlphaServer GS140 for
COMPAQ). The author also thanks the staff of The Supercomputer
Center(System A, B), Institute for Solid State Physics(ISSP),
University of Tokyo for the facilities provided by them.

\end{document}